\documentclass{iau}
\usepackage{graphicx}
\usepackage{natbib}
\usepackage{amsmath}
\usepackage{enumerate}

\title[eccentric mergers in galactic nuclei] 
{Gravitational-Wave Signatures of Highly Eccentric Stellar Binary Black-Holes in Galactic Nuclei}

\author[Evgeni Grishin, Isobel M. Romero-Shaw, Alessandro A. Trani]   
{Evgeni Grishin$^{1,2}$, Isobel M. Romero-Shaw$^3$, Alessandro A. Trani$^{4,5,6}$}
\affiliation{$^1$School of Physics and Astronomy, Monash University, Clayton, VIC 3800, Australia \\ 
$^2$OzGrav: Australian Research Council Centre of Excellence for Gravitational Wave Discovery, Clayton, VIC 3800, Australia \\email: {\tt evgeni.grishin@monash.edu}\\ $^3$H.H. Wills Physics Laboratory, Tyndall Avenue, Bristol BS8 1TL, United Kingdom \\$^{4}$Niels Bohr International Academy, Blegdamsvej 17, 2100 Copenhagen, Denmark\\$^{5}$INFN-Trieste, I-34127, Trieste, Italy\\
$^{6}$Departamento de Astronom\'ia, Facultad Ciencias F\'isicas y Matem\'aticas, Universidad de Concepci\'on, Avenida Esteban Iturra, Casilla 160-C, Concepci\'on, 4030000, Chile}

\pubyear{2025}
\volume{398}  
\pagerange{119--126}
\jname{Title of your IAU Symposium}
\editors{A.C. Editor, B.D. Editor \& C.E. Editor, eds.}
\begin{document}

\maketitle

\begin{abstract}
 A significant fraction of gravitational-wave mergers are expected to be eccentric in the Laser-Interferometer-Space-Antenna (LISA) frequency band, $10^{-4}-10^{-1}$\ Hz. Several LIGO–Virgo–KAGRA events show potential hints of residual eccentricity at $10$\ Hz, pointing to dynamical or triple origins for part of the population, where von-Zeipel-Lidov-Kozai oscillations can perturb both the eccentricity and the inclination of the binary. Moreover, the argument of pericentre, $\omega$, could be fully circulating, or librating, with a limited range for $\omega$. We use TSUNAMI, a regularised N-body code with 3.5PN corrections to identify four different orbital families: (i) circulating, (ii) small-amplitude and (iii) large-amplitude librating, and (iv) merging. We develop a new method to construct gravitational-wave waveforms using the quadrupole formula from the instantaneous acceleration in TSUNAMI. The four orbital families have distinct waveform phenomenologies, enabling them to be distinguished if observed in LISA. In particular, the properties of the tertiary companion can be inferred and serve as an independent mass measurement and distinguish field triple dynamics from galactic dynamics. 
\end{abstract}

Since the first discovery of GW150914, detections of binary compact object mergers via their gravitational-wave (GW) emission have revolutionised modern-day astronomy. While the LIGO-Virgo-KAGRA (LVK) detectors are sensitive to the $\sim10$---$10^4$~Hz band and capture the final few seconds of a binary black hole (BBH) merger, the sensitivity of the Laser Interferometer Space Antenna (LISA) is in the range $\sim10^{-4}\ \rm Hz$ to $\sim10^{-1}\ \rm Hz$, and will detect GWs from stellar-mass (and lower-mass) compact binaries several years before they merge (see \citealp{LISA2023} and references therein).   

The astrophysical origins of GW mergers are broadly divided into two categories: \textit{isolated}, where the binary evolves without interaction from external influences over its whole life, and \textit{dynamical}, where the binary is assembled by and influenced through gravitationally-driven interactions in densely-populated environments (such as globular and nuclear star clusters). The isolated scenario is expected to form mergers with quasi-circular orbits and small, close to aligned spins at $10$~Hz GW frequency, and their masses are expected to be limited by the pair-instability mass gap. Meanwhile, dynamical channels can result in eccentric, highly-spinning and spin-misaligned mergers at the same frequency, and their masses can be higher than the pair instability limit due to hierarchical mergers \citep[see][for details and references therein]{mandel22}.

\begin{figure}
    \centering
    \includegraphics[width=1.0\linewidth]{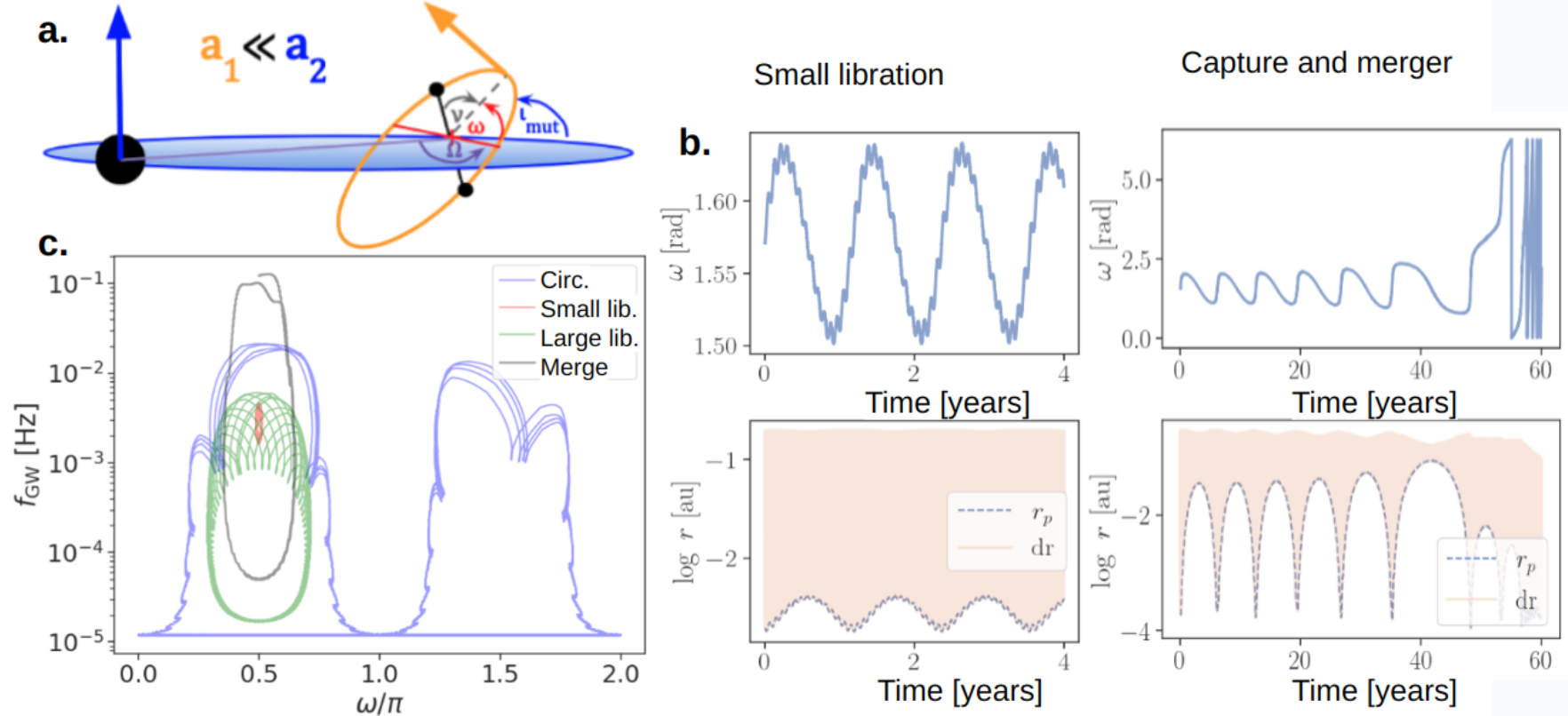}
    \caption{Examples of different evolution of a triple systems. a. Sketch of the triple system with semi-major axes $a_1\ll a_2$ and other orbital angles sketched for the inner orbit. b. Dynamical evolution of a system at small libration (left panels) and at an initially librating orbit that is decoupled and than captured due to GW emissions (right panels). c. The phase space of $f_{\rm GW} - \omega$ for the four different orbits.}
    \label{fig1}
\end{figure}

Isolated triples represent a middle ground between the isolated and dynamical evolution channels for BBH mergers. In particular, hierarchical triples can evolve as 'regular' binaries and only be affected by the tertiary on long, secular timescales. In particular, the inner eccentricity can be enhanced by the von-Zeipel-Lidov-Kozai (ZLK) mechanism (see \citealp{naoz16} for a review), which would accelerate the merger time and contribute to eccentric GW mergers. Therefore, mergers via isolated triples bear characteristics of both isolated and dynamical binaries: limited masses, but a wider distribution of spins than isolated binaries, and the potential for detectable eccentricity at $10$~Hz \citep[e.g.,][]{rod18}. Tentative evidence for eccentric mergers exists in the recent datasets, however the methods for complete inspiral-merger-ringdown eccentric waveform model approximants that distinguish eccentricity from other effects (e.g. spin precession) are still under development \citep{eccGW}.

For galactic nuclei, the perturbation of the binary stellar BH by the SMBH ZLK oscillations between the inner binary eccentricity $e_1$ and mutual inclination $\iota$ between the two orbits. The corresponding angle of the oscillation is the argument of pericenter $\omega$, which could either librate around an equilibrium value or complete a full circulation (see \ref{fig1} panel a. for a sketch). A hallmark of ZLK oscillations is that the peak eccentricity depends only on the initial inclination via $e_{\rm max} = \sqrt{1 - 5 \cos^2 \iota/3}$. However, this is true only when $\omega$ is circulating, while a librating solution will have somewhat limited eccentricity oscillations (e.g. \citealp{gri2024a}). 

\begin{figure}
    \centering
    \includegraphics[width=0.9\linewidth]{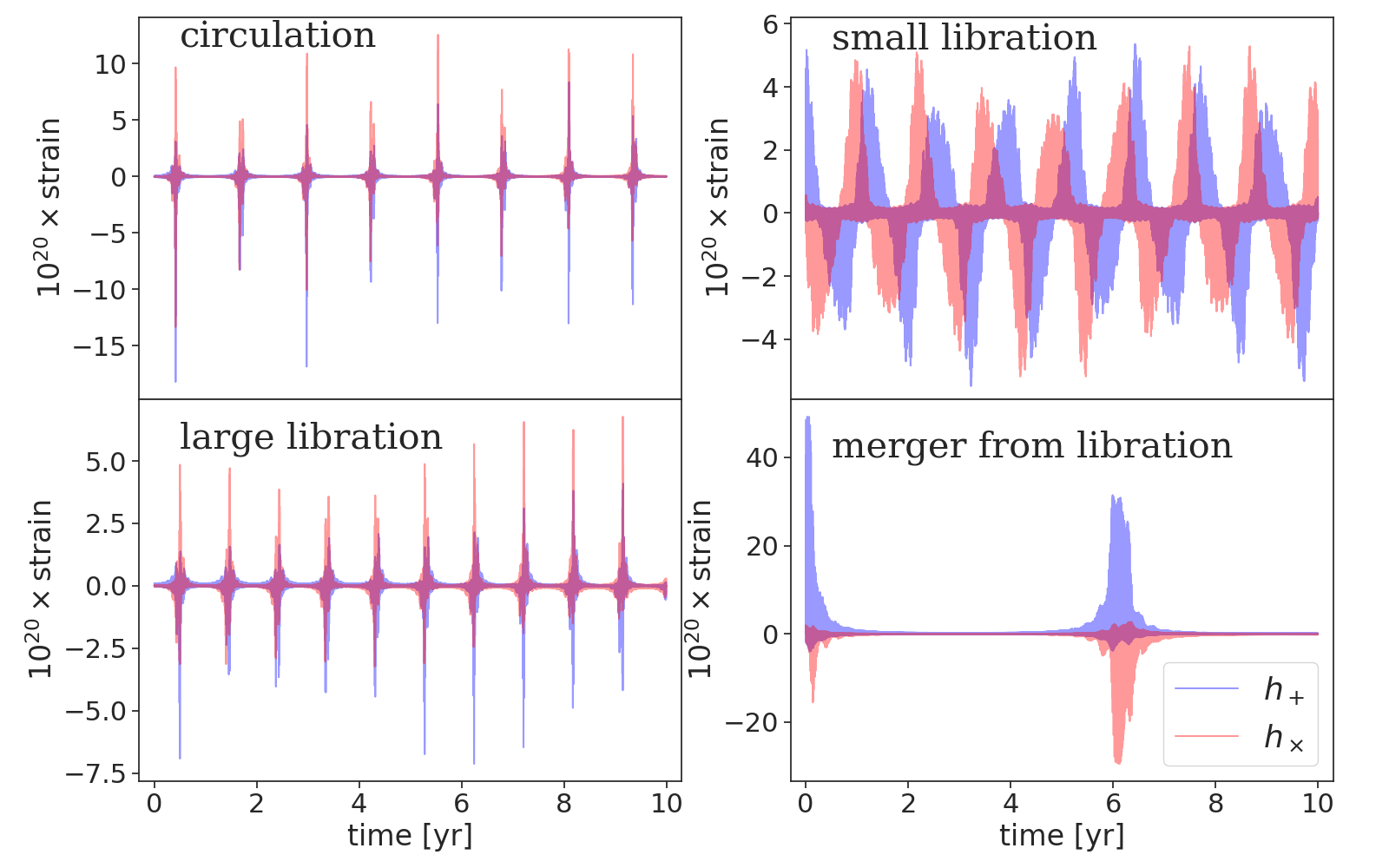}
    \caption{Empirical GW strain polarisations for the different orbits from panel c. of Fig. \ref{fig1}.}
    \label{fig2}
\end{figure}

We use the post-Newtonian code \texttt{TSUNAMI} \citep{tsunami-code}, a fast N-body code which implements chain-regularisation up to $3.5$ post-Newtonian (PN) order to integrate three body systems. We focus on four distinct classes of binary orbits: 

\begin{enumerate}[i]
    \item) Circulating orbits where $\omega$ completes a full revolution.
    \item) Librating orbits with small amplitude close to the fixed point $\omega_{\rm fix}=\pi/2$ and \citep{gri2024a}
\begin{align}    
    e_{\rm fix}=\sqrt{ 1 - \sqrt{ \frac{5(1+\frac{9}{8}\epsilon j_z)}{3(1-\frac{9}{8}\epsilon j_z)} } |j_z|}, \label{efix} \nonumber
\end{align}
where $j_z=\sqrt{1-e_1^2}\cos \iota_{\rm mut}$ is the 'Kozai constant' and $\epsilon \approx P_1/P_2$ is the period ratio and measured the strength of the double-averaging breakdown \citep{gri18}.
    \item) Librating orbits will large amplitude libration of $\omega$.
    \item) Orbits that are captured and merge during a highly eccentric encounter.
\end{enumerate}

The time evolution of orbits ii) and iv) is shown in panel b of Fig. \ref{fig1}, and the different GW signatures are shown in panel c. in the $f_{\rm GW}$-$\omega$ parameter space, where $f_{\rm GW}$
 \begin{align}
     f_{\rm GW} = \sqrt{\frac{G(m_1+m_2)}{a_1^3}}\frac{(1+e_1)^{1.1954}}{\pi (1-e_1^2)^{3/2}} \nonumber
 \end{align}
is the \cite{wen03} peak frequency for an eccentric orbit, and is a proxy for the orbital eccentricity. We see that the librating orbits (especially the one with small amplitude) are centred around $\omega=\pi/2$. 

In order to explore the GW signatures, we calculate the numerical GW strain into \texttt{TSUNAMI} from the quadrupole formula \citep{maggiore}, where we use the instantaneous accelerations computed directly from the code and include the effects of the tertiary and the PN terms.  In Fig. \ref{fig2} we plot the result of the two strain polarisations, $h_+$ and $h_\times$ for each of the four aforementioned orbits, but we need the same integration time of $10\ \rm yr$. The distance is $D=8\ \rm kpc$. We see that most of the orbits are `spiky' with sharp edges, except for the small libration which is smoother. Traditional waveform matching may fail in highly-eccentric orbits, and burst timing or wavelet transform may be required to detect these highly eccentric orbits \citep{knee2024}. 

\begin{figure}
    \centering
    \includegraphics[width=0.95\linewidth]{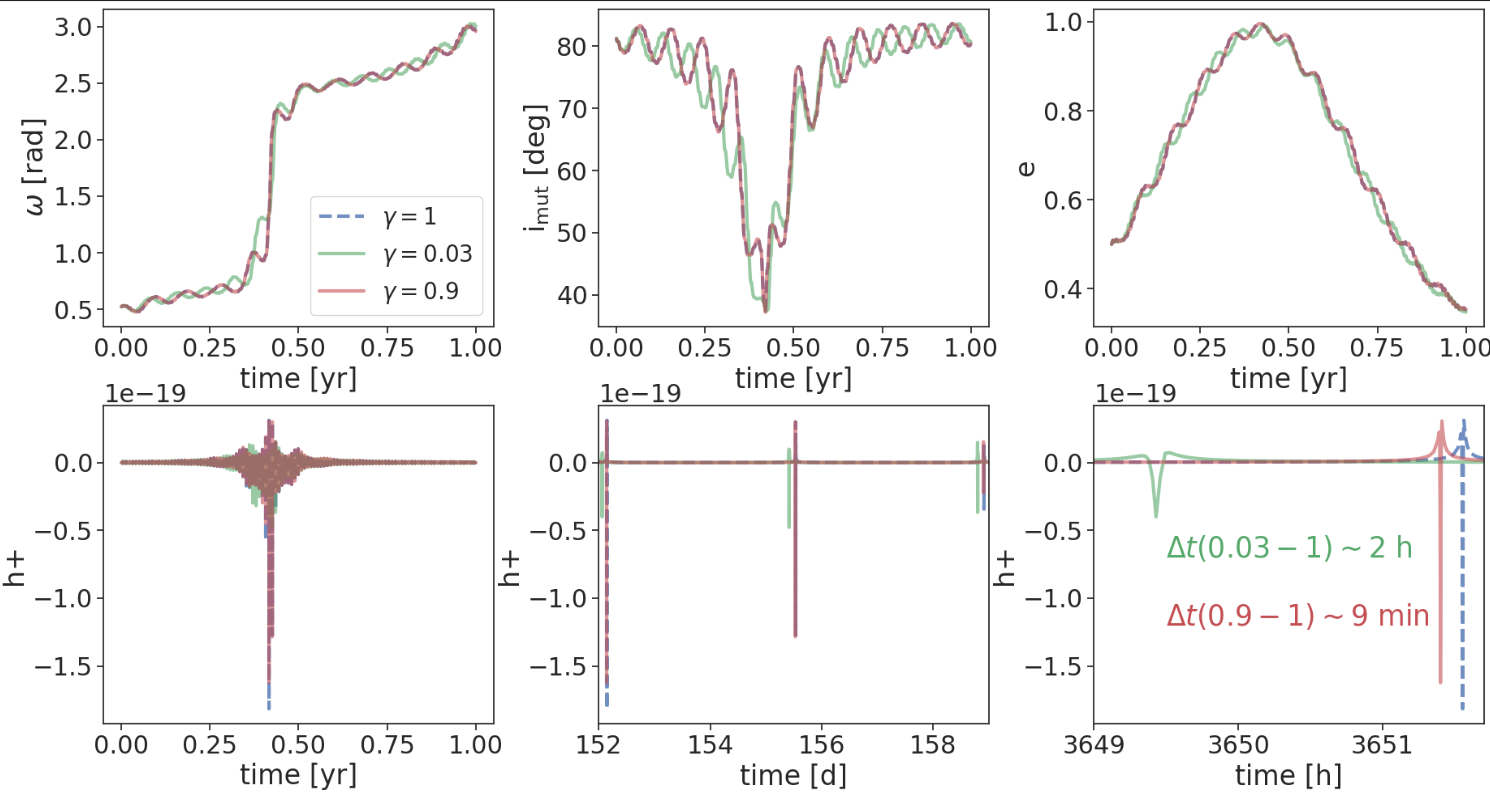}
    \caption{Circulating orbit for different mass and semi-major axis scaling. The SMBH mass is shifted by $m_3\to \gamma^3 m_3$ and the semi-major axis is shifted by $a_2\to \gamma a_2$, so $\gamma=0.03$ corresponds to $m_3=108 M_\odot$ a stellar mass BH. Top panel show the orbital evolution of the orbital elements. Bottom panel show the '+' strain polarisation, where the centre and right bottom panels are zoomed in versions of the strain data.}
    \label{fig3}
\end{figure}

Finally, we explore the prospects of independently measuring the tertiary mass. We rescale the system by $m_3\to\gamma^3m_3$ and $a_2\to \gamma a_2$ which keeps the secular timescale $t_{\rm sec} \propto a_2^3/m_3$ invariant. The top panels of Fig. \ref{fig3} show the dynamical evolution of the orbital elements for a slightly lower mass SMBH ($\gamma=0.9$, red curve) and a $\sim 100M_\odot$ stellar mass BH (green lines). We see that the evolution is almost identical for $\gamma=0.9$ since both systems are in the test particle limit. However, the field triple case $\gamma=0.03$ is noticeably different due to the feedback onto the outer orbit. 
From the bottom panels of Fig. \ref{fig3}, we see that the GW strain peak is associated with the peak of the eccentricity. Upon closer zoom in it is evident that the differences in the arrival times $\Delta t$  (2 hours and $9$ minutes for $\gamma=0.03,0.9$ respectively) are much longer than the width of the peaks, suggesting that unmodeled burst timing methods are more suitable for detecting these orbits.
 
\textit{In summary}, we identify different orbital families depending on the behaviour of the argument of pericentre $\omega$ of the inner binary. We build a tool to extract empirical gravitational-wave strain polarisations from the N-body simulations and analyse the different GW signatures. Most of the waveforms are bursty due to the high eccentricities involved, except for the small libration which is more regular. Keeping the secular timescale intact, even a small change in the tertiary mass and separation will lead noticeable changes in the time of arrival of the GW bursts. Unmodeled burst models can be efficient for this problem, but it suffers from higher dimensionality due to the large parameter space of triple systems and its chaotic evolution.

\vspace{-0.6cm}
\begin{thebibliography}{}

\bibitem[Amaro-Seoane, \etal\ (2023),]{LISA2023}{Amaro-Seoane P., \etal\ 2023} \textit{Living Reviews in Relativity}, 26, 2

\bibitem[Grishin\ (2024),]{gri2024a}{Grishin E.,\ 2024} \textit{MNRAS}, 533, 486

\bibitem[Grishin \etal\ (2018),]{gri18}{Grishin E., Perets, H. B., Fragione G.,\ 2018} \textit{MNRAS}, 481, 4907

\bibitem[Knee \etal\ (2024),]{knee2024}{Knee A. M., McIver J., Naoz S., Romero-Shaw I. M., Hoang B.-M., Grishin E.,\ 2024} \textit{ApJL}, 971, L38

\bibitem[Maggiore (2007),]{maggiore}{Maggiore M., 2007} \textit{Gravitational Waves: Volume 1:
Theory and Experiments}, Oxford University Press, ISBN: 9780198570745

\bibitem[Mandel \& Broekgaarden (2022),]{mandel22}{Mandel I., Broekgaarden F. S., 2022} \textit{Living Reviews in Relativity}, 25, 1

\bibitem[Morras \etal\ (2025),]{eccGW}{Morras G., Pratten G., Schmidt P.,\ 2025} \textit{ArXiv}, 2503.15393

\bibitem[Naoz (2016),]{naoz16}{Naoz, S.,\ 2016} \textit{ARAA}, 54, 441

\bibitem[Rodriguez \& Antonini (2018),]{rod18}{Rodriguez, C. L., Antonini F.,\ 2018} \textit{ApJ}, 863, 7

\bibitem[Trani \& Spera (2023),]{tsunami-code}{Trani A. A., Spera M.,\ 2023} \textit{The Predictive Power of Computational Astrophysics as a Discover Tool}, 362, 404

\bibitem[Wen (2003),]{wen03}{Wen L.,\ 2018} \textit{ApJ}, 598, 419

\end{thebibliography}
\end{document}